# Observation of Magnetic-Anisotropy Crossover and High-Temperature Skyrmions in the Dirac Magnet $Fe_3Ge$ with a Distorted Kagome Lattice

Yalei Huang[1,12], Xiaowei Lv[2,12], Bin Li[3], Chunqiang Xu[4,*], Dhanarajagopal Alltrin[5], Xiaoxuan Ma[6], Wanting Yang[6], Wei Zhou[7,*], Xiangzhuo Xing[8], Wen-Chin Lin[9], Raman Sankar[5], Michael Smidman[10], Shixun Cao[6], Dong Qian[11], Renchao Che[2,*] and Xiaofeng Xu[1,*]

[1]Zhejiang Provincial Key Laboratory of Quantum Precision Measurement, School of Physics and Optical Engineering, Zhejiang University of Technology, Hangzhou 310023, China

[2]Laboratory of Advanced Materials, Shanghai Key Lab of Molecular Catalysis and Innovative Materials, State Key Laboratory of Coatings for Advanced Equipment, College of Smart Materials and Future Energy, Fudan University, Shanghai 200438, China

[3]School of science, Nanjing University of Posts and Telecommunications, Nanjing 210023, China

[4]School of Physical Science and Technology, Ningbo University, Ningbo 315211, China

[5]Institute of Physics, Academia Sinica, Nankang, Taipei, R.O.C. 11529 Taiwan

[6]Materials Genome Institute, Shanghai University, Shanghai 200444, China

[7]School of Electronic and Information Engineering, Suzhou University of Technology, Changshu 215500, China

[8]Laboratory of High Pressure Physics and Material Science (HPPMS), School of Physics and Physical Engineering, Qufu Normal University, Qufu 273165, China

[9]Department of Physics, National Taiwan Normal University, Taipei 116, Taiwan

[10]School of Physics, Zhejiang University, Hangzhou, 310058, China

[11]Key Laboratory of Artificial Structures and Quantum Control (Ministry of Education), School of Physics and Astronomy, Shanghai Jiao Tong University, Shanghai 200240, China

[12]These authors contributed equally: Yalei Huang, Xiaowei Lv.

[*]Corresponding authors: Chunqiang Xu, Wei Zhou, Renchao Che, Xiaofeng Xu.

**Email:** xuchunqiang@nbu.edu.cn, wei.zhou@cslg.edu.cn, rcche@fudan.edu.cn, xuxiaofeng@zjut.edu.cn

## Abstract

Topological materials that simultaneously host robust high-temperature skyrmions and nontrivial electronic band structures have attracted tremendous interest owing to their distinctive advantages for both fundamental research and prospective technological applications. Here, we report the observation of robust skyrmions in the Dirac kagome magnet $Fe_3Ge$, which exhibits a high Curie temperature of ~ 650 K. At room temperature, $Fe_3Ge$ shows a large intrinsic anomalous Hall conductivity of ~ 380 $\Omega^{-1}cm^{-1}$, originating from its nontrivial electronic band topology. Systematic magnetization measurements reveal a spin reorientation transition at ~ 375 K, indicating a crossover from easy-plane to easy-axis magnetic anisotropy. Below the spin reorientation temperature, a large topological Hall effect is observed, arising from microscopic noncoplanar spin structures. Lorentz transmission electron microscopy shows that mesoscopic skyrmions are stabilized in the easy-axis magnetic anisotropy regime and persist over an exceptionally wide temperature window of 375-650 K, far exceeding that of most previously reported skyrmion-hosting materials. These results establish $Fe_3Ge$ as a promising platform for exploring diverse topological properties, with strong potential for advancing future high-temperature spintronic applications, ranging from next-generation information storage to logic computing devices.

## Introduction

Magnetic skyrmions are particle-like spiral spin textures with nontrivial topology, and have emerged as a focus of condensed matter physics and spintronics research, owing to their intrinsic topological stability and immense potential for information storage and processing [1-5]. These textures are responsible for diverse phenomena, such as the topological Hall effect (THE) [6], skyrmion magnetic resonance [7], and thermally induced ratchet motion [8]. Their nanoscale dimensions, particle-like robustness, and exceptionally low current thresholds for motion make them promising candidates for next-generation memory and logic devices with high density and low energy consumption [9-11]. Early observations of skyrmions were predominantly reported in non-centrosymmetric chiral magnets, such as MnSi [3,6], MnGe [12] and $Cu_2OSeO_3$ [13], where the Dzyaloshinskii-Moriya interaction (DMI) plays a crucial role in their stabilization. However, such systems generally exhibit Curie temperatures ($T_C$) below room temperature and have narrow temperature windows for skyrmion formation, severely constraining their applications in devices. In contrast, centrosymmetric magnets with easy-axis magnetic anisotropy (EAMA) have emerged as an alternative platform for hosting skyrmions [4,14-17]. In such systems, the competition between dipolar interactions, magnetic anisotropy, and ferromagnetic exchange can stabilize skyrmions without invoking the DMI. Representative examples include MnNiGa [15], $Fe_3Sn_2$ [16], and $Mn_5Ge_3$ [18], where skyrmions persist over considerably broader temperature ranges compared with their DMI-stabilized counterparts. These advances suggest that centrosymmetric magnets provide a promising route toward realizing thermally stable skyrmions.

Frustrated magnetic systems, particularly kagome magnets, provide a versatile platform for stabilizing high-temperature skyrmions. Theoretically, magnetic frustration combined with EAMA has been demonstrated to promote skyrmionic spin textures and significantly suppress thermal fluctuations [19,20]. Experimentally, this paradigm is exemplified by the kagome ferromagnet $Fe_3Sn_2$[16,21], which hosts skyrmions over a wide temperature range with the EAMA and a high $T_C$ of approximately 630 K. Despite these advances, magnetic materials that simultaneously combine a high $T_C$, a broad skyrmion stability window, and tunable magnetic anisotropy remain exceedingly rare.

Beyond real-space topological spin textures, topological band structures in momentum space can also generate unconventional electronic transport phenomena, such as the large intrinsic anomalous Hall effect (AHE) [22] and anomalous Nernst effect (ANE) [23]. Importantly, topological properties in real space and momentum space are not mutually exclusive and may coexist within a single material [18]. However, magnetic systems that simultaneously host high-temperature skyrmions and nontrivial topological electronic band structures in momentum space remain relatively scarce. Such systems provide a unique platform for exploring diverse emergent topological phenomena and advancing the development of multifunctional spintronic devices.

A particularly promising candidate in this context is hexagonal $Fe_3Ge$, which crystallizes in a $D0_{19}$ type structure (space group $P6_3/mmc$) [24,25]. The projection of the Fe sublattice onto the basal plane can be visualized as a triangular lattice formed by a twisted triangular tube of face-sharing octahedra (Fig. 1a and b). As shown in Fig. 1c, the Fe atoms form a 2D kagome lattice within the (0001) plane (i.e., the *ab* plane), with the central Ge ion slightly displaced from its geometric center. The two distinct nearest-neighbor Fe-Fe bond lengths are 2.672 Å and 2.497 Å, resulting in a distorted kagome lattice [26]. Notably, this material exhibits a high $T_C$ of 630-680 K [25,27]. At high temperatures, it displays EAMA and undergoes a spin reorientation transition at $T_{SR}$ ~ 350-380 K [25,28]. Recent studies verified $Fe_3Ge$ to be a topological semimetal with Dirac fermions [26,29]. The large intrinsic AHE and ANE were observed, originating from the Berry curvature concentrated near the nontrivial bands in momentum space [26]. In addition, a pronounced THE and topological Nernst effect (TNE) have been observed in $Fe_3Ge$ below $T_{SR}$, which are attributed to finite scalar spin chirality associated with microscopic noncoplanar spin structures[26,30]. These intriguing properties motivate a deeper investigation into the evolution of magnetic anisotropy and the potential emergence of skyrmions in $Fe_3Ge$, particularly in the high-temperature regime above the spin reorientation transition.

In this work, we report the observation of skyrmions in $Fe_3Ge$ under EAMA. Magnetization measurements reveal a spin reorientation transition at $T_{SR}$ ~375 K, where the magnetic anisotropy changes from easy-plane at low temperatures to easy-axis at high temperatures. $Fe_3Ge$ exhibits a large intrinsic AHE originating from its nontrivial electronic band structures. In addition, a sizable THE is observed below $T_{SR}$, originating from microscopic noncoplanar spin structures. With increasing temperature, Lorentz transmission electron microscopy (LTEM) measurements demonstrate that above $T_{SR}$, where the system exhibits EAMA, skyrmions can be stabilized and persist up to the $T_C$ ~ 650 K. This observation establishes an exceptionally broad skyrmion stability window, far exceeding that of most previously reported skyrmion-hosting materials. These findings demonstrate that $Fe_3Ge$ can serve as a versatile platform for investigating various topological properties, while also offering broad prospects for advancing future spintronic applications.

## Results and Discussion

### Characterizations of $Fe_3Ge$ crystals

The $Fe_3Ge$ single crystals investigated in this work were synthesized by the Sn-flux method (details given in the Materials and Methods). Fig. 1d presents the energy-dispersive X-ray spectroscopy (EDX) spectrum of the as-grown single crystal. The atomic ratio of Fe to Ge is determined to be 2.97:1, in excellent agreement with the ideal 3:1 stoichiometry. As shown in Fig. 1e, EDX elemental mapping images confirm a uniform distribution of Fe and Ge throughout the crystal, verifying the absence of elemental segregation during the synthesis process. Detailed single-crystal X-ray characterizations confirm the high quality of the grown crystals (Fig. S1 and Table S1 and S2). To identify the crystallographic orientations of the $Fe_3Ge$ single crystals, Fig. 1f and g display the Laue and X-ray diffraction (XRD) patterns from the $(01\bar{1}0)$ plane, respectively. In addition, the clear and bright diffraction spots of the Laue pattern attest to the high quality of the as-grown single crystals. To further validate the crystal structure of the sample, high-angle annular dark-field scanning transmission electron microscopy (HAADF-STEM) was performed, as shown in Fig. 1h. The inset of Fig. 1h shows that the experimental HAADF-STEM image along the [0001] zone axis matches well with the corresponding structural model, clearly highlighting its distorted kagome lattices. Collectively, these characterizations confirm both the correct stoichiometry and high crystallinity of the as-grown $Fe_3Ge$ single crystals used in this study. For clarity, we define $[2\bar{1}\bar{1}0]$, $[01\bar{1}0]$ and [0001] as *x*-, *y*-, and *z*-axes, respectively, as illustrated in the Fig. 1i.

**Magneto-transport properties and anisotropic magnetoresistance effect**

We further fabricated a Hall device as shown in the inset of Fig. 2a and conducted magneto-transport measurements. Fig. 2a shows the zero-field longitudinal resistivity ($\rho$-$T$) of $Fe_3Ge$ with the current applied along the *c*-axis ($I \parallel c$), which exhibits good metallicity characterized by a residual resistivity ratio (RRR) of ~ 18. In the low-temperature (low-$T$) regime, the $\rho$-$T$ curve can be well described by $\rho = \rho_0 + aT^{2.5}$, where $\rho_0$ denotes the residual resistivity. The $T^{2.5}$ dependence, previously observed in other ferromagnets [31-33], is generally attributed to the combined contributions from electron-electron, electron-magnon and electron-phonon scattering. The field-dependent Hall resistivity ($\rho_{xz}$-$\mu_0H$) was then measured at various temperatures, as shown in Fig. 2b. All $\rho_{xz}$-$\mu_0H$ curves exhibit pronounced nonlinearity, the hallmark of the anomalous Hall effect. Remarkably, the anomalous Hall conductivity (AHC) reaches a magnitude of $\sigma_{xz}^{A} \sim 380\ \Omega^{-1}\mathrm{cm}^{-1}$ at 300 K (Fig. S2), a value comparable to, or even surpasses, those reported for typical topological kagome materials, such as $Mn_3Sn$[22], $Mn_3Ge$[34], and $Fe_3Sn_2$[35]. Furthermore, both the scaling analysis and the theoretical calculations (Fig. S2) suggest that the AHE in $Fe_3Ge$ is dominated by the intrinsic mechanism, which is consistent with prior reports [26,36,37]. This behavior underscores the significant contribution of Berry curvature arising from massive Dirac gaps in momentum space. Furthermore, the topological Hall effect (THE) was extracted by subtracting the contributions of the ordinary Hall effect and the AHE (Fig. 2b and Fig. S3). As detailed below, this large THE originates from microscopic noncoplanar spin structures with finite scalar spin chirality.

Fig. 2c and 2d present the field dependence of transverse and longitudinal magnetoresistance (MR) at various temperatures, respectively. For the transverse configuration ($\mu_0H \parallel y$, $I \parallel z$), $Fe_3Ge$ exhibits a negative MR over a broad temperature range (100 K ≤ $T$ ≤ 300 K), which can be attributed to the suppression of spin-dependent scattering by the applied field [38,39]. In the low-$T$ regime, a positive MR emerges under high magnetic fields, originating from the quenching of fluctuating magnetic moments combined with the conventional orbital MR due to the Lorentz effect [40]. Notably, as shown in Fig. 2d, the longitudinal MR ($\mu_0 H \parallel I \parallel z$) first increases with the applied field and then decreases, resulting in a distinct peak at a characteristic field of $\mu_0 H_p$. This peak gradually shifts to higher fields with decreasing temperature. Such peak-like MR features have been previously observed in magnetic materials and are generally attributed to the anisotropic magnetoresistance (AMR) effect [41]. In ferromagnets with easy-plane magnetic anisotropy (EPMA), spins can be canted away from the easy *ab*-plane when an external magnetic field is applied along the hard axis (*c*-axis). This spin reorientation modifies the relative orientation between the current and magnetization from perpendicular (at zero field) to nearly parallel (above the saturation field), thereby increasing the resistivity and yielding a positive MR contribution. As discussed later, $Fe_3Ge$ exhibits EPMA at $T$ < 375 K. Moreover, the peak position $\mu_0 H_p$ in the MR is consistent with the saturation field ($\mu_0 H_s$) extracted from isothermal magnetization (*M-H*) curves (Fig. S4). To eliminate uncertainties associated with the demagnetization factor, the MR and *M-H* curves were measured on samples with identical geometric dimensions. These results demonstrate that the peak-like feature in the longitudinal MR of $Fe_3Ge$ originates from the AMR effect. The electronic band structure calculations reveal distinct impacts of magnetic anisotropy on the electronic properties (Fig. 2e-g and Fig. S5). When Fe moments are aligned along the $x$ and $y$ directions, the corresponding band structures are remarkably similar, indicating comparable electronic environments for these in-plane (IP) magnetization directions. In contrast, when the Fe moments are oriented along the $z$ axis, i.e., out-of-plane (OOP) direction, the two Dirac fermions respond differently: the Dirac gap at point 1 is significantly enhanced, whereas that at point 2 is only weakly affected. Meanwhile, the bands near the Fermi level along the M-K path become notably flattened. In addition, along the Γ-A-L-H-A path, the bands evolve from a degenerate state to a split configuration, reflecting the lifting of degeneracy induced by out-of-plane magnetization. These band structure modifications underscore the strong magnetic anisotropy in $Fe_3Ge$, where the OOP magnetization significantly alters the electronic structure compared to the IP orientations. Moreover, we calculated the $\sigma_{zx}^{A}$ for different spin configurations in $Fe_3Ge$, as shown in Fig. S6. The magnitude of $\sigma_{zx}^{A}$ is considerably larger when the Fe moments are aligned along the $y$ axis than when they are oriented along the $z$ axis, consistent with previous reports [27]. These results indicate that the spin configuration has a significant impact on the transport properties of $Fe_3Ge$.

**Magnetic properties and magnetic-anisotropy crossover**

To investigate the intrinsic magnetic properties of $Fe_3Ge$ single crystals, temperature-dependent magnetization (*M-T*) measurements were carried out under $\mu_0 H$ = 0.1 T, with the field oriented in both OOP ($\mu_0 H \parallel z$) and IP ($\mu_0 H \parallel y$) configurations (Fig. 3a). Owing to the in-plane magnetic isotropy of $Fe_3Ge$ (Fig. S7), the *y* direction is chosen as a representative in-plane orientation throughout this work. A clear downturn is observed at $T_{SR}$ ~ 375 K, consistent with the temperature of the spin-reorientation transition. Subsequently, isothermal magnetization *M-H* curves were measured for both orientations, as depicted in Fig. 3b and 3c. The effect of the demagnetizing field was accounted for using the relation $\mu_0 H_{int} = \mu_0 H - 4\pi N M_v$, where $M_v$ denotes the magnetization per unit volume and the prefactor $4\pi$ is required for the cgs units. Based on the sample's geometric dimensions, the demagnetization factor was estimated to be $N_y$ = 0.42 ($\mu_0 H \parallel y$) and $Nz$ = 0.10 ($\mu_0 H \parallel z$). The negligible coercive field ($\mu_0 H_c$) across the entire temperature range confirms that $Fe_3Ge$ exhibits the behavior of a soft ferromagnet. This conclusion is further corroborated by the magneto-optical Kerr effect (MOKE) measurements, which yield $\mu_0 H_c$ values of only a few tens of Oersteds at 300 K (Fig. 3d).

As shown in Fig. 3e and Fig. S8, the OOP *M-H* curves consistently exhibit a lower $\mu_0 H_s$ when $T > T_{SR}$, revealing EAMA. Upon cooling below $T_{SR}$, the IP *M-H* curves display a lower $\mu_0 H_s$, characteristic of EPMA. These results unambiguously demonstrate the occurrence of a spin reorientation transition at $T_{SR}$ in $Fe_3Ge$. Moreover, the anisotropy field ($\mu_0 H_k$) is defined as the critical field at which the difference in magnetization between the two field orientations drops below 2%. Fig. 3f presents the temperature dependence of $\mu_0 H_k$, which is consistent with prior reports[28]. To further quantify the magnetic anisotropy, the effective magnetic anisotropy constant $K_{eff}$ was calculated using the equation [42,43]: $K_{eff} = \mu_0 \int_0^{M_s} [H_c(M) - H_{ab}(M)] dM$, where $M_s$ denotes the saturation magnetization, $H_{ab}$ and $H_c$ represent IP and OOP fields, respectively. A positive (negative) $K_{eff}$ corresponds to EAMA (EPMA). The temperature-dependent $K_{eff}$ is presented in Fig. 3f. For $T < T_{SR}$, $K_{eff}$ is clearly negative; $K_{eff} \sim -4.1 \times 10^5$ J m$^{-3}$ at $T$ = 2 K. The absolute value of $K_{eff}$ decreases with increasing temperature, signifying a gradual weakening of EPMA. As shown in Fig. 3g, $Fe_3Ge$ undergoes a spin reorientation transition at $T_{SR}$, where the magnetic anisotropy reverses; above $T_{SR}$, $K_{eff}$ switches from negative to positive. At $T$ = 400 K, $K_{eff}$ ~ 0.36 × $10^5$ J $m^{-3}$, confirming the presence of EAMA, a property of great significance for practical applications. By accounting for the evolution of magnetic anisotropy, the physical origin of the recently observed large THE [26,30] and TNE [26] below $T_{SR}$ becomes more transparent. At low temperatures, the large absolute value of $K_{eff}$ indicates that the spins of $Fe_3Ge$ are predominantly aligned within the *ab* plane. Consequently, the scalar spin chirality, defined as $\boldsymbol{S}_i\cdot(\boldsymbol{S}_j\times\boldsymbol{S}_k)$, is strongly suppressed, resulting in weak THE and TNE. As the temperature increases, the absolute value of $K_{eff}$ gradually decreases, reflecting a weakened EPMA and a reduced energy barrier between the easy-plane and easy-axis orientations. Under these circumstances, spin non-coplanarity becomes energetically more favorable, inducing a progressive enhancement of the THE with increasing temperature.

Fig. 3h compares the $K_{eff}$ of $Fe_3Ge$ with that of various other ferromagnetic materials. Compared with other reference magnets [16,18,40,44-46], $Fe_3Ge$ demonstrates remarkable tunability; its magnetic anisotropy can be switched from EPMA to EAMA, exhibiting a broad adjustable range. It is well established that magnetic anisotropy plays a crucial role in the formation and stabilization of spin structures. Our results suggest that the spin structures in $Fe_3Ge$ can undergo significant variations across a wide temperature range, driven by the changes in magnetic anisotropy. In this context, $Fe_3Ge$ serves as a promising platform for exploring the evolution of spin structures over an extended temperature range.

**Bloch-type skyrmions and magnetic phase diagram of $Fe_3Ge$**

To directly visualize the magnetic structures in $Fe_3Ge$, LTEM was performed on a [0001]-oriented square lamella with an approximate thickness of 100 nm and a width of 2 μm. The lamella used for LTEM was prepared by the focused ion beam (FIB) milling technique. Fig. 4a-d and Fig. S9 illustrate the magnetic domain structures of $Fe_3Ge$ in zero magnetic field. When $T < T_{SR}$, in-plane domains are observed, where domain walls (white or dark lines in Fig. 4a and 4b) intersect at the boundaries between multiple domains. Furthermore, through transport of intensity equation (TIE) analysis, magnetic induction maps acquired at 300 K and 370 K further reveal the existence of spin vortices and antivortices located at domain boundaries (highlighted by white and yellow dashed circles, respectively). As the temperature increases ($T_{SR} < T <$ 650 K), labyrinthine domains emerge as the ground states at zero magnetic field (Fig. 4c); this domain configuration arises from the competition between magnetic anisotropy, dipole interactions, and ferromagnetic exchange interactions. The alternating black and bright stripes correspond to magnetic moments oriented upward and downward, respectively, revealing that the magnetization of the domains in the flake sample prefers to be aligned along the *z*-axes, namely, EAMA. This behavior is consistent with that observed in bulk magnets at $T > T_{SR}$. Note that the sample was not tilted during characterization, enabling us to identify the domain wall as Bloch-type rather than Néel-type based on their different imaging conditions. Upon further heating to 650 K, as shown in Fig. 4d, the magnetic domain contrast vanishes, indicating the $T_C$ of $Fe_3Ge$.

To systematically investigate the magnetic structures, experiments were conducted under varying magnetic fields within the temperature range of $T_C > T > T_{SR}$. Fig. 4e and Fig. S10 display the evolution of magnetic structures under magnetic fields at 420 K. When a magnetic field is applied downward along the *z* axis, the magnetic stripe domains gradually shrink with increasing field. When $\mu_0 H$ = 727 mT, the edges of certain stripe domains start to detach, forming isolated spin textures. At 997 mT in Fig. 4e, except for dumbbell-shaped worm domains, these isolated spin textures can be categorized into three distinct types based on their LTEM contrasts, as highlighted by red, green, and blue dashed circles (labeled 1, 2, and 3). The spin textures enclosed by the red and green circles exhibit concentric bright-dark rings corresponding to Bloch-type skyrmions with clockwise (CW, helicity $\gamma = -\pi/2$) and counterclockwise (CCW, helicity $\gamma = \pi/2$) spin rotations, respectively. In contrast, the spin texture marked by the blue dashed circle, shows a dark-bright-dark-bright contrast, characteristic of a topologically trivial bubble. Moreover, through TIE analysis, the in-plane magnetization distributions of both skyrmions and trivial bubbles were resolved, as illustrated in Fig. 4f.

The coexistence of skyrmions with opposite helicities is a hallmark of dipolar skyrmions [47]. In $Fe_3Ge$, skyrmions with helicities of $-\pi/2$ and $\pi/2$ are observed with equal population, providing strong evidence for their dipolar origin. Although the distorted kagome lattice in $Fe_3Ge$ may allow a weak local DMI, such a weak DMI is expected to lift the helicity degeneracy and induce an imbalance between the two helicities (Fig. S11) [48], which is not observed experimentally. Moreover, the typical skyrmion diameter $d_{sk}$ is measured to be 250 nm, which further supports that the skyrmion stabilization is dominated by magnetic dipolar interactions [4]. Collectively, these results demonstrate that $Fe_3Ge$ hosts dipolar skyrmions alongside trivial magnetic bubbles.

As the magnetic field gradually increases, the size of the skyrmions decreases slightly, whereas that of the trivial bubbles reduces significantly. At fields above $\mu_0H$ ~ 0.1 T, only skyrmions remain in the images. Skyrmions, characterized by a nontrivial topological charge of $Q = \pm1$, are stabilized by an energy barrier that inhibits their continuous deformation into spin structures with different topological charges. In contrast, trivial bubbles, with $Q = 0$, are topologically equivalent to the fully aligned spin state. It is this topological distinction that accounts for why skyrmions can withstand higher perpendicular magnetic fields (and thus higher Zeeman energy) than trivial bubbles [48]. When the magnetic field approaches 1200 mT, all spin textures disappear, indicating that the sample reaches ferromagnetic saturation, with all spins aligned along the *z* axis.

A series of LTEM measurements enabled us to outline the evolution of spin textures across a broad range of temperatures and magnetic fields (Fig. S12 and S13), thereby constructing a comprehensive magnetic phase diagram. As illustrated in Fig. 5a, under an OOP magnetic field ($\mu_0H \parallel z$), $Fe_3Ge$ exhibits a sequence of magnetic states upon varying temperature and field. For $T < T_{SR}$, vortices and antivortices are stabilized below $\mu_0H_s$. With increasing temperature, a spin reorientation transition occurs at $T_{SR}$, accompanied by a switch in magnetic anisotropy from EPMA to EAMA. In this regime, as the magnetic field increases, stripe domains gradually transform into skyrmions via an intermediate mixed state. Once the temperature exceeds $T_C$, the long-range magnetic order collapses, and the system enters the paramagnetic phase. In the temperature range above $T_{SR}$, THE and TNE signals may in principle be present. However, their origin would no longer be attributed to microscopic noncoplanar spin structures. Instead, mesoscopic skyrmions proliferate in this regime, as directly observed in this work, and may contribute to the THE and TNE. While dipolar-stabilized skyrmions are relatively large in size and thus not anticipated to produce a measurable bulk transport signal, they may nonetheless yield a finite real-space contribution to topological transport when probed in reduced-geometry samples. Extending transport measurements to such high temperatures and thin-sample configurations remains experimentally challenging. Remarkably, the skyrmion phase maintains excellent stability over an extensive temperature range of 375-650 K, which is a key feature demonstrating its potential for high-temperature spintronic applications. Furthermore, prior studies have demonstrated that chemical substitution can effectively lower the spin reorientation temperature, enabling EAMA already at room temperature [28,49]. This strategy offers a promising pathway for the realization of room-temperature skyrmions in $Fe_3Ge$-based systems.

For comparison, Fig. 5b summarizes the reported skyrmion temperature windows across various materials. In contrast to other centrosymmetric and noncentrosymmetric materials [2,6,13,15,16,18,45,50-52], $Fe_3Ge$ not only hosts skyrmions over an exceptionally broad temperature interval but also exhibits one of the highest $T_C$ among all skyrmion-hosting materials. LTEM observations further confirm that once nucleated, skyrmions in $Fe_3Ge$ are highly robust. Although mixed textures (skyrmions and trivial bubbles) are observed in the present lamellae, it should be noted that these dipolar-stabilized skyrmions are highly sensitive to extrinsic geometric parameters. Further optimization of skyrmion size, density, and nucleation behavior can be achieved through precise control of sample thickness and geometric confinement (a detailed discussion is provided in the Supplementary Materials) [53,54]. This approach represents a promising direction that warrants further investigation. The realization of stable skyrmions at such elevated temperatures is crucial for future technological applications in magnetic storage and spintronic devices. In addition, Table S4 provides a comparative summary of magnetic materials that support both skyrmions and nontrivial electronic band structures [16,18,35,40,50,54-61]. This comparison reveals that, in addition to its high Curie temperature and broad skyrmion stability window, $Fe_3Ge$ also exhibits a relatively large anomalous Hall response among this category of materials. Therefore, $Fe_3Ge$ emerges as a promising platform not only for exploring diverse topological phenomena but also for advancing the development of next-generation spintronic devices.

## Conclusion

In this work, we have successfully synthesized high-quality single crystals of kagome Dirac magnet $Fe_3Ge$ and systematically elucidated its distinctive transport and magnetic properties. We demonstrate that $Fe_3Ge$ exhibits a large intrinsic anomalous Hall effect originating from its nontrivial electronic band structures. Comprehensive magnetic characterization reveals a spin reorientation transition at approximately $T_{SR}$ ~ 375 K, accompanied by a crossover in magnetic anisotropy from an easy-plane to an easy-axis configuration. The constructed magnetic phase diagram reveals the transitions among various magnetic structures as functions of temperature and magnetic field. Notably, robust skyrmions are stabilized over an exceptionally broad temperature range of 375-650 K in the easy-axis magnetic anisotropy regime, far exceeding the stability windows of most previously reported skyrmions systems. As a prospective avenue for future exploration, the spin reorientation temperature might be further tuned via chemical substitution, thereby enabling the realization of room-temperature skyrmions in this kagome system. Our results naturally account for the large topological Hall and topological Nernst effects recently observed in this material below the spin reorientation temperature. Overall, these findings establish $Fe_3Ge$ as a versatile platform for investigating both topological electronic band structures in reciprocal space and topological spin textures in real space, while also offering promising opportunities for advancing future spintronic applications.

## Methods

### Single Crystal Growth

High-quality $Fe_3Ge$ single crystals were synthesized via the Sn-flux method. High-purity Fe powders (99.99%), Ge lumps (99.9999%), and Sn powders (99.99%) in a molar ratio of 1:1:20 were loaded into an alumina crucible and sealed in a quartz tube under high vacuum. The tube was heated to 1080 °C and held at this temperature for 10 h, followed by slow cooling to 700 °C at 3 °C $h^{-1}$. Excess molten flux was removed by centrifugation to isolate the single crystals, which exhibited shiny surfaces and a hexagonal rod-shaped morphology.

### Sample Characterizations

The chemical compositions were characterized by EDX (HGSTFlexSEM-1000). XRD measurements were performed by using the powder X-ray diffractometer (Bruker D2 PHASER) and single-crystal X-ray diffractometer (Rigaku XtalLAB Synergy). Laue diffraction pattern was recorded with a back-reflection Laue detector (Try-SE. Co., Ltd.). Atomic-scale HAADF-STEM images were performed on a FEI Titan$^3$ Themis G3 60–300.

### Electrical Transport and Magnetization Measurements

Gold wires were used as electrodes of the bulk devices, and they were connected by silver epoxy. (Magneto-)resistance and the Hall resistance were measured using the standard four-probe method in the Cryogenic system. By reversing the field polarities, the even signal in field was defined as MR and the odd component was calculated as the Hall resistivity. Magnetization measurements were conducted using a superconducting quantum interference device magnetometer (MPMS, Quantum Design). Magneto-optical Kerr effect was measured using a magneto-optical Kerr microscope (Evico Magnetics GmbH), with the measurement setup illustrated in Fig. S14.

### LTEM observation and image simulation

For the LTEM observation, a [0001]-oriented $Fe_3Ge$ lamella was prepared via FIB Ga+ ion milling (Helios Nanolab 600I; FEI) using standard lift-out procedures. The magnetic structure of the sample was characterized by a modified Lorentz transmission electron microscopy (JEM-2100F) at an acceleration voltage of 200 KV. The high-resolution in-plane magnetic induction maps were processed by QPT software based on the transport of intensity equation. The LTEM images were simulated by using the MALTS code.

**Theoretical calculation**

Electronic structure calculations were performed using the full-potential linearized augmented plane wave (FPLAPW) method as implemented in the WIEN2K package [62]. The generalized gradient approximation with the Perdew–Burke–Ernzerhof (PBE) functional was employed for the exchange-correlation potential [63]. Muffin-tin radii were set to 2.0 a.u. for both Fe and Ge atoms. Self-consistent field calculations utilized a 9×9×10 *k*-point mesh for Brillouin zone sampling, while Fermi surface construction employed a denser 37×37×39 *k*-point mesh to ensure adequate resolution. To calculate the anomalous Hall conductivity, we constructed a tight-binding model based on maximally localized Wannier functions [64] using Fe *s*, *d* and Ge *p* orbitals. The Wannierization procedure was carried out using Wien2wannier [65] , and subsequent AHC calculations were performed with WannierTool [66]. Spin-orbit coupling effects were included in all calculations unless explicitly stated otherwise.

## Data availability

The data that support the findings of this study are available from the corresponding authors upon reasonable request.

## Acknowledgments

This work was supported by National Natural Science Foundation of China (Grants No. 12274369, No. 12304071, and No. 12374116) and the China Postdoctoral Science Foundation (Grants No. 2024M751931 and No. 2025M773352). A portion of work was supported by Zhejiang Provincial Natural Science Foundation of China (Grant No. LZ25A040003 and No. LQN26A040014) and Fundamental Research Project JG-WL-2025057. D.Q. was financially supported by the National Key R&D Program of China No. 2021YFA1400100. R.S. acknowledges the financial support provided by the Ministry of Science and Technology in Taiwan under Project No. NSTC-114-2124-M-001-003 and No. NSTC-114-2112M001-045-MY3, as well as support from Academia Sinica for the budget of AS-

iMATE11412.

## Author contributions

Y.H. and X.L. contributed equally to this work. Y.H. and X.X. conceived the project and designed the experiments; C.X. synthesized the sample; C.X., X.M., W.Y., S.C. and D.A., W. L., R. S., M.S. carried out the basic characterizations; C.X. and W.Z. performed the magnetization measurements; X.L. and R.C. conducted the LTEM and HAADF-STEM experiments; B.L. carried out the first-principles calculations; Y.H. and X.X. analyzed data and wrote the manuscript. All authors contributed to the discussion of the results and the improvement of the manuscript.

## Competing interests

The authors declare no competing interests.

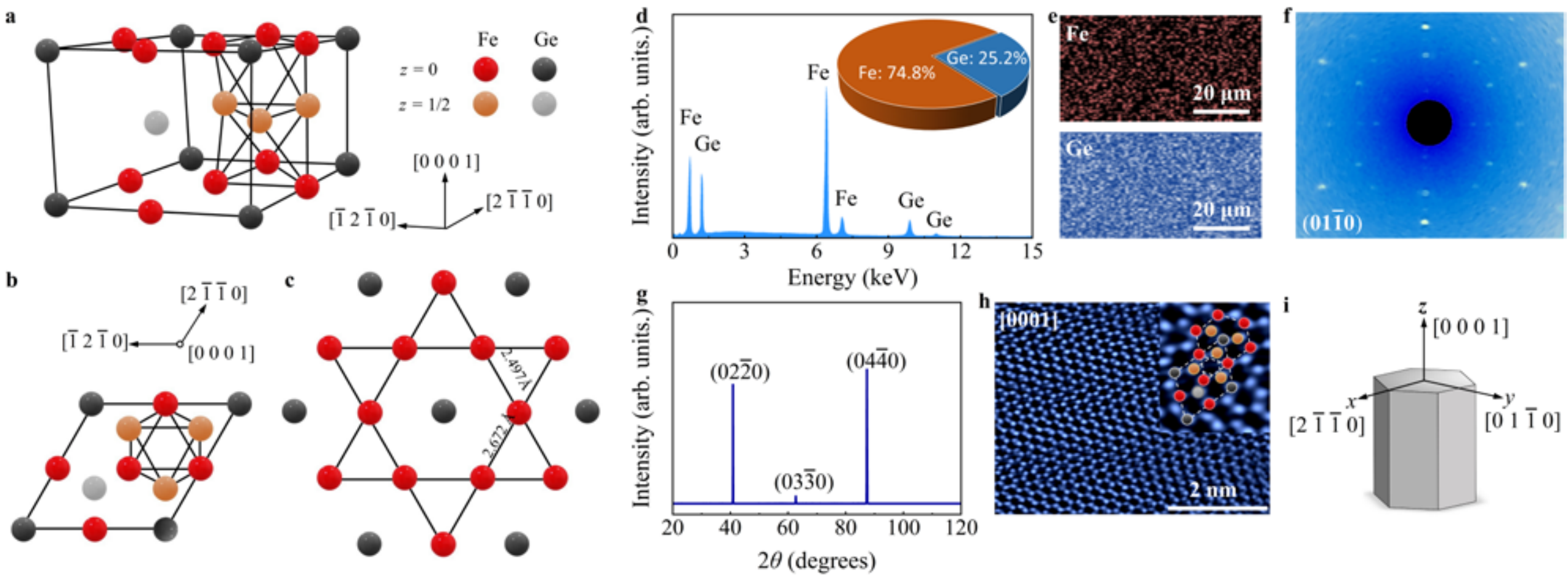

**Figure 1. Crystal structures and characterizations of $Fe_3Ge$. a** Crystallographic unit cell of $Fe_3Ge$. Fe and Ge occupy single crystallographic sites, whereas atoms in the $z = 0$ and $z = 1/2$ planes are colored differently for clarity. **b** Top view of the unit cell along the *c*-axis projected onto the *ab* plane. **c** An individual *ab* plane of $Fe_3Ge$. Here, the directions $[2\bar{1}\bar{1}0]$, $[\bar{1}2\bar{1}0]$ and [0001] correspond to the *a*-, *b*- and *c*-axes, respectively. **d** Representative EDX spectrum collected from an as-grown $Fe_3Ge$ single crystal. The inset shows the measured element molar (atomic) ratio. **e** EDX mapping of Fe and Ge for the as-grown single crystal revealing the spatial distribution and homogeneity of the constituent elements. **f** Laue pattern of a $Fe_3Ge$ single crystal recorded along the $(10\bar{1}0)$ direction. **g** XRD pattern of an as-grown $Fe_3Ge$ single crystal showing the $(01\bar{1}0)$ diffraction peaks. **h** HAADF-STEM image of a $Fe_3Ge$ single crystal acquired along the [0001] zone axis. The upper inset displays a magnified view, and the observed atomic arrangement is consistent with the structural model. **i** Schematic illustration showing our definition of the $[2\bar{1}\bar{1}0]$, $[01\bar{1}0]$ and [0001] directions as the *x*-, *y*-, and *z*-axes, respectively.

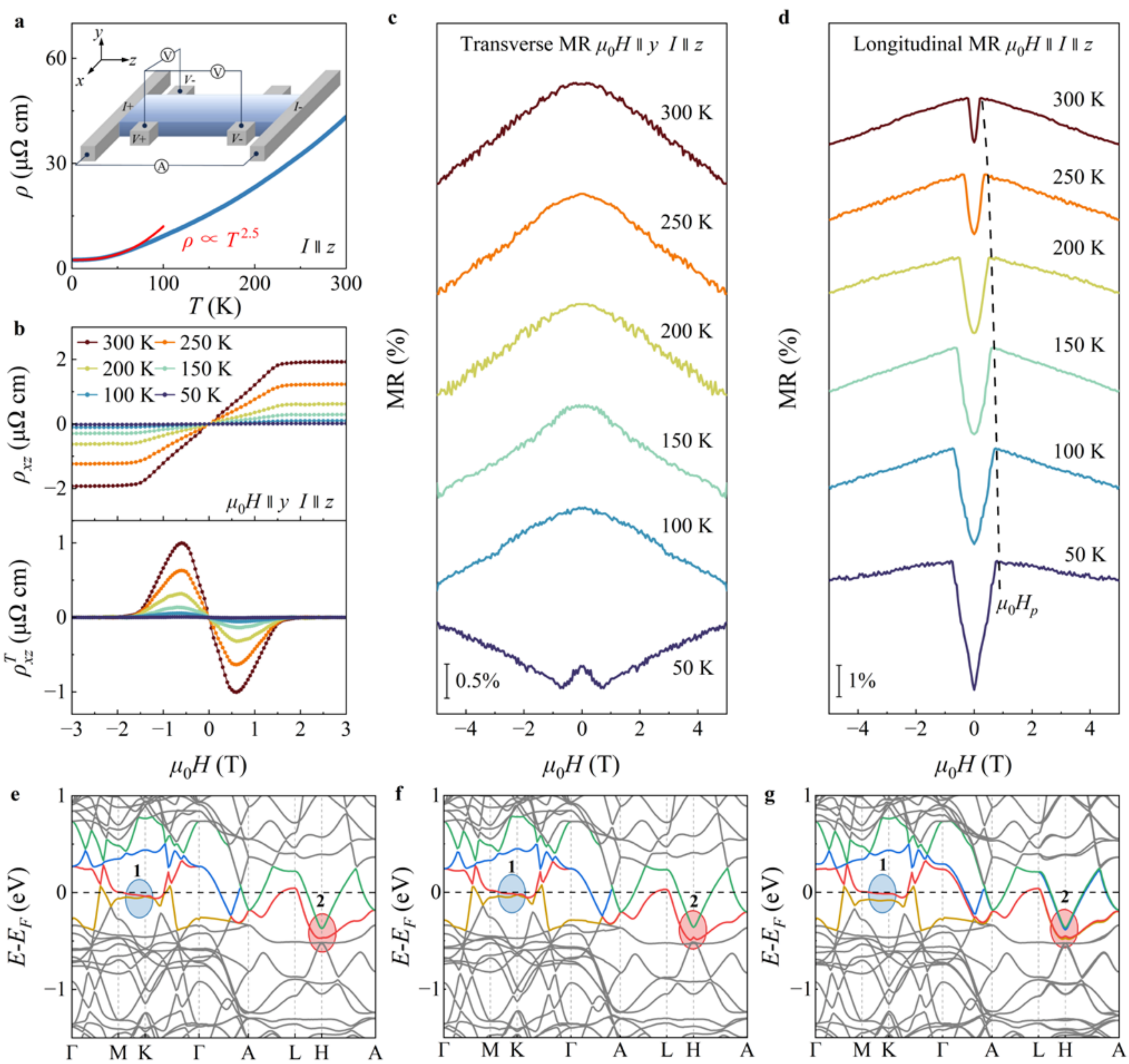

**Figure 2. Magneto-transport measurements and band structures of $Fe_3Ge$ single crystals. a** Temperature-dependent longitudinal resistivity with the current applied along the *z*-axes. The red solid line represents a fit to $\rho = \rho_0 + aT^{2.5}$. The inset shows the schematic of device configuration for the electrical transport measurements. **b** Magnetic field dependence of Hall resistivity $\rho_{xz}$ (top row) and topological Hall resistivity $\rho^{\mathrm{T}}_{xz}$ (bottom row) at various temperatures. **c** Transverse MR with $\mu_0 H \perp I \parallel z$ at various temperatures. **d** Longitudinal MR with $\mu_0 H \parallel I \parallel z$ at various temperatures. **e-f** Calculated electronic band structures of $Fe_3Ge$ for the Fe moments aligned along the *x*-, *y*-, and *z*-axes, respectively. Four bands crossing the Fermi level are highlighted in different colors. Two sets of Dirac points below $E_F$ are highlighted by the blue (1) and red (2) circles.

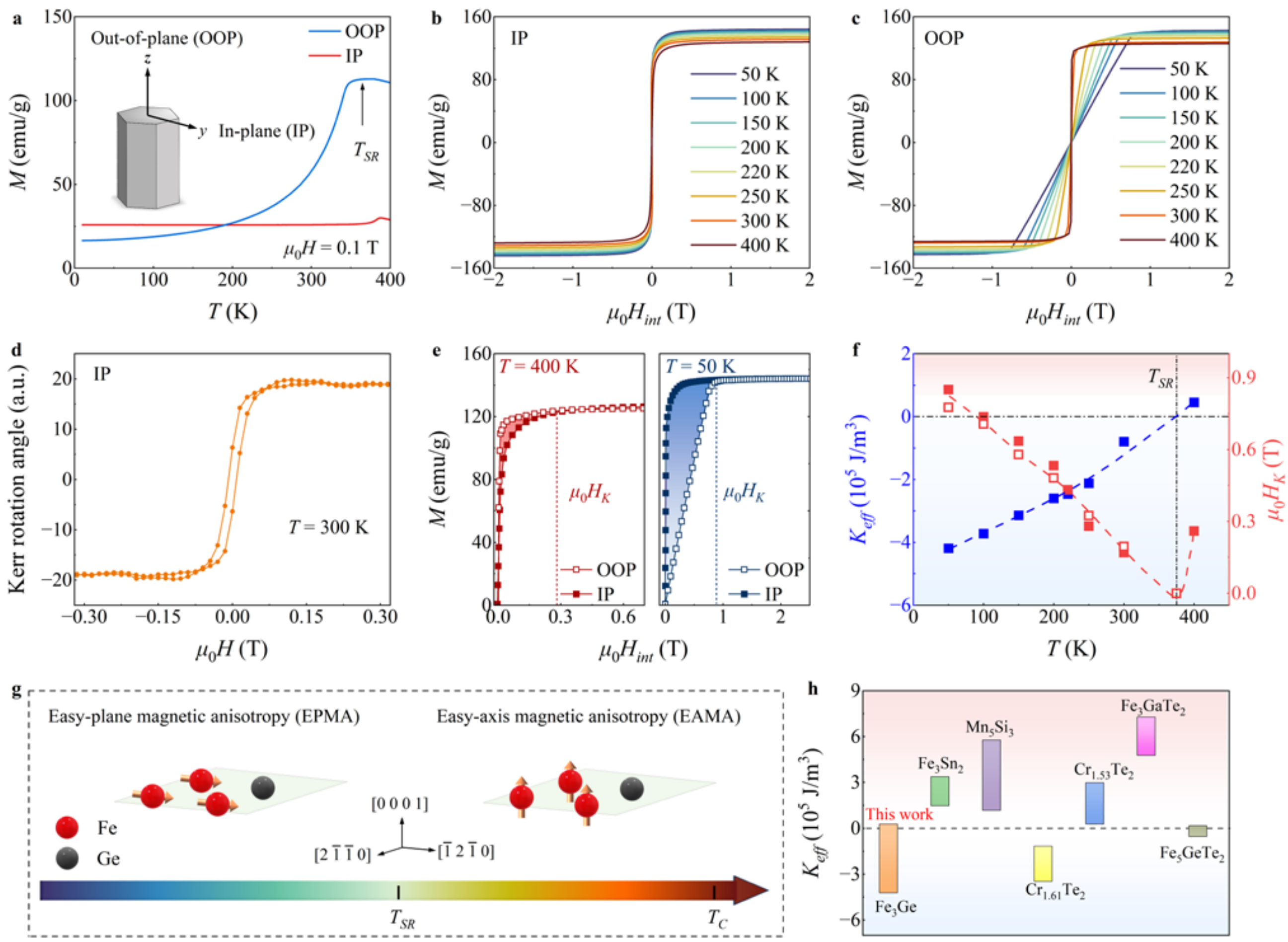

**Figure 3. Magnetic properties of $Fe_3Ge$ single crystals. a** Temperature-dependent magnetization under an applied field of 0.1T for both OOP and IP fields. The inset shows the corresponding magnetic field configurations. **b, c** Field-dependent magnetization at various temperatures for OOP and IP field configurations, respectively. **d** Kerr rotation hysteresis loops at 300 K. **e** Enlarged views of OOP and IP *M-H* curves measured at 400 K and 50 K, respectively. **f** Temperature dependence of magnetic anisotropy energy density $K_{eff}$ and anisotropy field $\mu_0 H_k$. For $\mu_0 H_k$, red solid squares correspond to results from this work, whereas the red open squares represent values reported previously[28]. The dashed lines serve as a guide to the eye. **g** Schematic magnetic phase diagram and spin structures of $Fe_3Ge$, illustrating the evolution of magnetic anisotropy with temperature. **h** Comparison of the $K_{eff}$ for various ferromagnetic materials[16,18,40,44-46].

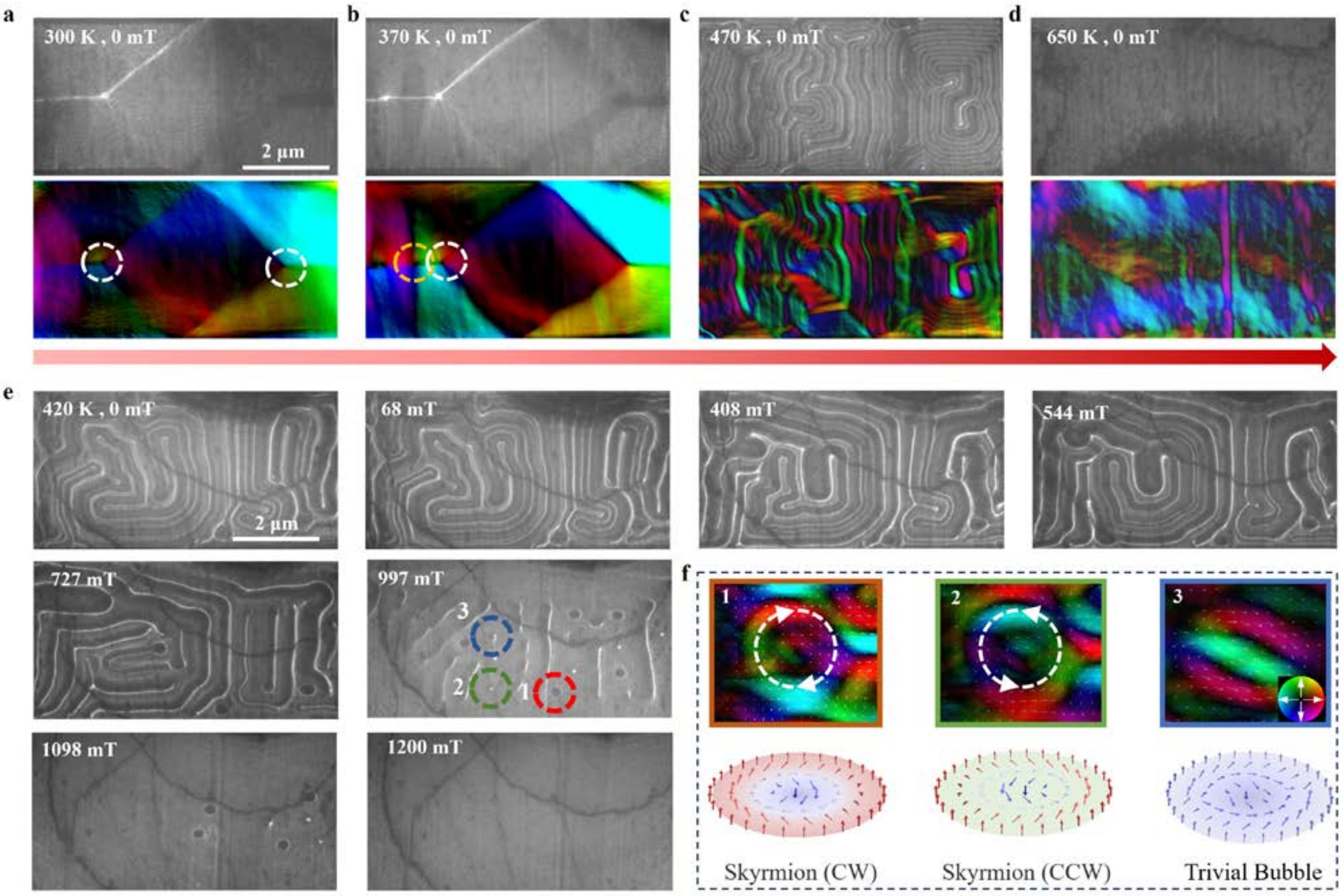

**Figure 4. Spin textures of $Fe_3Ge$ single crystals. a-d** LTEM images (top row) and the corresponding magnetic induction maps (bottom row) of a $Fe_3Ge$ lamella recorded under zero magnetic field at 300 K, 370 K, 470 K, and 650 K. Magnetic vortices and antivortices are highlighted by white and yellow dashed circles, respectively. The imaging plane corresponds to the crystallographic *ab* plane. **e** Over-focused LTEM images of spin textures in $Fe_3Ge$ at 420 K under various magnetic fields. Red and green dashed circles denote Bloch-type skyrmions with opposite helicity, while the blue dashed circle marks a trivial magnetic bubble. **f** Color-coded magnetic induction maps (top row) of the three spin textures marked in **e**, together with schematic illustrations of their spin configurations (bottom row). White arrows correlated the colors of the induction map with the orientation of the magnetization.

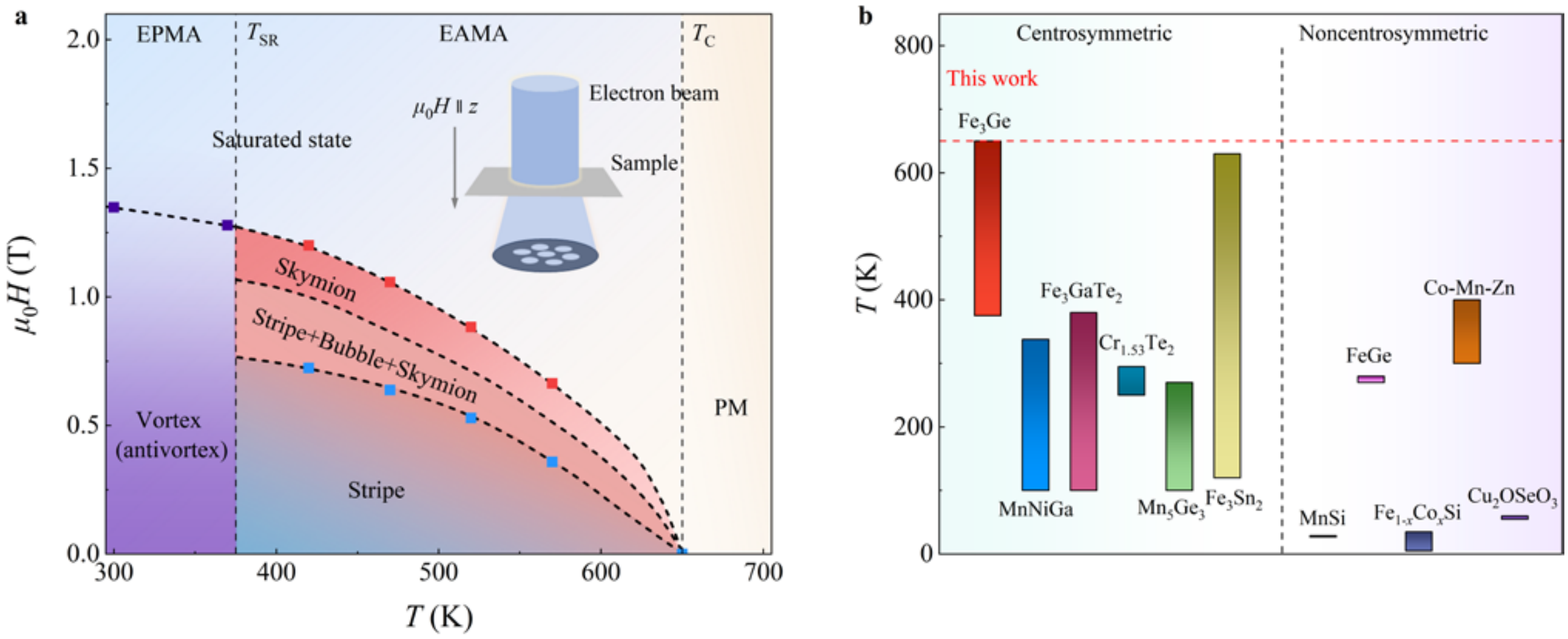

**Figure 5. Magnetic phase diagram. a** Magnetic phase diagram of $Fe_3Ge$ sketching the evolution of spin textures as a function of temperature and magnetic field ($\mu_0 H \parallel z$). EPMA: easy-plane magnetic anisotropy; EAMA: easy-axis magnetic anisotropy. The inset illustrates the LTEM setup. **b** Comparison of the temperature stability ranges of skyrmions in various materials [2,6,13,15,16,18,45,50-52].